# The Role of Boolean Function in Fractal Formation and it's Application to CDMA Wireless Communication

Somnath Mukherjee
Dr. B. C. Roy Engineering College
Email: somnath.7.mukherjee@gmail.com
Durgapur, West Bengal-713206, INDIA

Pabitra Kumar Ghosh
Dr. B. C. Roy Engineering College
Email: pabitra06114@gmail.com.
Durgapur, West Bengal-713206, INDIA

*Abstract*— In this paper, a new transformation is generated from a three variable Boolean function 3, which is used to produce a self-similar fractal pattern of dimension 1.58. This very fractal pattern is used to reconstruct the whole structural position of resources in wireless CDMA network. This reconstruction minimizes the number of resources in the network and so naturally network consumption costs are getting reduced. Now -a -days resource controlling and cost minimization are still a severe problem in wireless CDMA network. To overcome this problem fractal pattern produced in our research provides a complete solution of structural position of resources in this Wireless CDMA Network.

**Keywords-** *Boolean functions, Level of Boolean function, Fractal Pattern, Wireless CDMA Networks, BTS, Wireless Network Port*.

## I. Introduction

A three variable Boolean function is used to generate the fractal pattern for the implementation of Wireless CDMA Network. More precisely, one transformation is generated from the three variable Boolean function named as 3 (named according to the Wolfram naming conventions [2]) which can produce the self-similar fractal pattern in a significant manner of increasing the level of Boolean function. In papers [1], [2] we have explored an application in the formation of self-similar and chaotic fractal formations, using one transformation named as 'Carry Value Transformation' (CVT). In this paper, we have explored the algebraic beauties of the newly generated function and their application towards wireless communication problem basically on Wireless Code-Division-Multiple-Access (CDMA) network. It is to be noted that we have generated the fractals using a computational program written in C language and the compiler version is Borland C -3.0 by the newly defined transformation from the Boolean function 3.

At first, we have generated one square matrix using the defined transformation as obtained in [1], and then we got the fractal using the matrix entities. The Wireless CDMA Network is designed based on the above generated fractal pattern. The recourses of the wireless CDMA Network if placed in the fractal pattern then the efficiency and effectiveness would be enhanced on the basis of design and maintenance cost. Interestingly, the services into this designed network are uniformly distributed.

## II. Review of Earlier Works and Fundamental Concepts

In paper [3], resource allocation for the purpose of energy efficiency has been explored, but in this paper with the help of fractal geometry we implement the designing of the positional structure of resources in wireless CDMA network.

Let us first warm up ourselves with some fundamentals, which are related to the current paper.

*A. Boolean Function:*

A Boolean function f $(x_1, x_2, x_3, x_4 \ldots \ldots x_n)$ variables is defined as a mapping from $\{0,1\}^n$ into $\{0,1\}$. It is also interpreted as the output column of its truth table f which is a binary string of length $2^n$. For n -variables the number of Boolean functions is $2^{2^n}$ and each Boolean function is denoted as $f_R^n$ known as the function number R (also interpreted as rule number R), of n -variable. Here R is the decimal equivalent of the binary sequence (starting from bottom to top, with top is the LSB) of the function in the Truth Table, and numbering scheme is proposed by Wolfram and popularly known as Wolframs naming convention.

*B. Fractal Pattern*

The pattern, which reserves a fractional real number, as its fractal dimension is known as Fractal pattern. Here we use the similarity dimension as a fractal dimension. The similarity dimension is defined as follows
For a self-similar pattern, there is a relation between the scaling factor 'S' and the number of pieces 'N' into which the pattern can be divided and that relation is

$N = 1/S^D$ this relation can be equivalently written
$\qquad D = \log N/\log (1/S)$.
This 'D' is called fractional dimension or fractal dimension (self-similarity dimension).

## C. BTS (Base Transceiver Station)

A base transceiver station or cell site (BTS) is a piece of equipment that facilitates wireless communication between user equipment (UE) and a network. UEs are devices like mobile phones (handsets), WLL phones, computers with wireless internet connectivity, WiFi and WiMAX gadgets etc. The network can be that of any of the wireless communication technologies like GSM, CDMA, WLL, WAN, WiFi, WiMAX etc. BTS is also referred to as the radio base station (RBS), node B (in 3G Networks) or, simply, the base station (BS).

## D. Wireless Network Port

Wireless Network ports are the points that contains the device to emit web to provide users uninterrupted network, it may be compared with the BTS(Base Transceiver Station ) of the mobile communication

## E. Why. Wireless CDMA Network

There are several wireless networks like TDMA(Time Division Multiple Access), FDMA(Frequency Division Multiple Access),CDMA(Code Division Multiple Access).In FDMA the available bandwidth is divided into frequency bands, that means each station is allocated to send its data. In TDMA each station share the bandwidth of the channel in time , each station is allocated a time slot during which it can send data. In both (FDMA & TDMA) case a switching technique is required to provide the network service to the station. CDMA(Code Division multiple Access) Network  is a third-generation (3G) wireless communications and it is a form of multiplexing, which allows numerous signals to occupy a single transmission channel, optimizing the use of available bandwidth. The technology is used in ultra-high-frequency (UHF) cellular telephone systems in the range from 800-MHz to 1.9-GHz. In CDMA one channel carries all transmissions simultaneously where channel means a common path between sender and receiver [4]. CDMA simply means communications with different codes. Let us assume we have three stations namely station-1, station-2 and station-3 are connected with the same channel The data from station-1 is d1, from station-2 is d2 and so on are allocated for each station.  Each station have to be assigned also a  particular code e.g. c1,c2 and so on. The data carried out by the channel is the sum of  the terms(d1.c1,d2.c2,d3.c3…), that means the data in the channel at any instant time is the sum of the values of d1.c1,d2.c2,d3.c3 and so. Any station wants to receive the data that is sent from the other station ,have to multiply  the data on the channel by the code of the sender. That means here is no time delay for transmission of data like any others wireless communication. In case of CDMA there is no such switching technique is required to provide the network service. So the implementation of CDMA wireless network by the model of fractal pattern  is possible in practically to provide the network services  among  the  stations.

## F. Carry Value transformation (CVT)

In [1] we have defined a new transformation named as CVT and shown its use in the formation of fractals.

If $a = (a_n, a_{n-1}, ..., a_1)$ and $b = (b_n, b_{n-1}, ..., b_1)$ are two n-bit strings then $CVT(a,b) = (a_n \wedge b_n, a_{n-1} \wedge b_{n-1}, ..., a_1 \wedge b_1, 0)$ is an (n+1) bit string, belonging to the set of non-negative integers, and can be computed bit wise by logical AND operation followed by a 0.

Conceptually, CVT in binary number system is same as performing the bit wise XOR operation of the operands (ignoring the carry-in of each stage from the previous stage) and simultaneously the bit wise ANDing of the operands to get a string of carry-bits, the latter string is padded with a '0' on the right to signify that there is no carry-in to the LSB (the overflow bit of this ANDing being always '0' is simply ignored).

Example:
Consider the CVT of the numbers $(13)_{10} \equiv (1101)_2$ and $(14)_{10} \equiv (1110)_2$. Both are 4-bit numbers. The carry value is computed as follows:

```
Carry:  1 1 0 0 0
Augend:   1 1 0 1
Addend:   1 1 1 0
XOR:      0 0 1 1
```

Carry genereted in ith column saved in (i-1)th column

In the above example, bit wise XOR gives $(0011)_2 \equiv (3)_{10}$ and bit wise ANDing followed by zero-padding gives $(11000)_2 \equiv (24)_{10}$. Thus $CVT(1101,1110) = 11000$ and equivalently in decimal notation one can write $CVT(13,14) = 24$. In the next section, a  new notion of CVT named as *Level Sensitive Carry Value Transformation* is discussed.

III Level Sensitive Carry Value Transformation

The Level Sensitive Carry Value Transformation is defined on the domain (ZxZxN) and it maps to  Z. e. In other words, LSCVT is a mapping from ZxZxN->Z where  $Z$  is set of non-negative integers and N is the set of all natural number .
Here we have considered Boolean function 3 and firstly the the decimal number '3' is converted to it's binary form then the corresponding
binary values are assigned according to the functional values of the truth table of a three variable Boolean function.

For Boolean function 3 the functional values of the truth table are assigned to the  corresponding  binary values of  3.

| Function | | Value |
|---|---|---|
| f(0,0,0) | ------ | 1. |
| f(0,0,1) | ------ | 1. |
| f (0,1,0) | ------ | 0. |
| f(0,1,1) | ------ | 0. |
| f (1,0,0) | ------ | 0. |
| f(1,0,1) | ------ | 0. |
| f(1,1,0) | ------ | 0. |
| f(1,1,1) | ------ | 0. |

Now the binary values of the corresponding any number in a matrix are

$Z_1$ ----- the binary form of $Z_1$
$Z_2$ ----- the binary form of $Z_2$
$N$ ----- the binary form of $N$
------------------------------------------
The corresponding functional value from the above truth table.

Example:-
$Z_1=4, Z_2=5, N=4$
Then the corresponding binary values are
```
    4  ----- 1 0 0
    5  ----- 1 0 1
    4  ----- 1 0 0
         ---------------
              0 1 0
```
So the decimal value of the corresponding binary value is 2.
Thus LSCVT(100,101,100) = 010 and equivalently in decimal notation it can be written LSCVT(4,5,4) = 2.

IV. Generation of Self-Similar Fractal Pattern Using LSCVT

A matrix is constructed that contains only the carry values (or even terms) defined above between all possible integers a's, b's and c's are arranged in an ascending order of x, y and z-axis respectively. We observe some interesting patterns in the matrix. We would like to make it clear how the matrix is constructed

**Step 1**: Arrange all integers 0,1,2,3,4 …(as long as we want) in ascending order and place it in three axis(x,y,z) in the matrix.

**Step 2:** Compute LSCVT (a,b,c) for c =1,2,3…using Boolean function 3 of three variable and store it in decimal form at (a,b,c) position.

**Step 3:** Then we notice on the pattern of value '0' and we have made it a specific color (green) and the other values except '0' also made a specific color (red) in the matrix. The pattern made by '0' in the matrix shown as a fractal (describe in figure -1below).

.

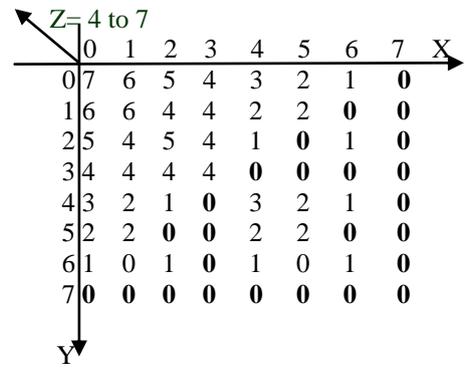

Figure-1     Fractal formation in a matrix

Here Y represents the row of the matrix, X represents the Column and Z represents the level of the matrix.
For an example, the positional decimal value of the matrix in the position (6,4 ,4) is 1.
For Boolean function 3 the functional values of the truth table are assigned to the corresponding binary values of 3.

| Function | | Value |
|---|---|---|
| f(0,0,0) | ------ | 1. |
| f(0,0,1) | ------ | 1. |
| f (0,1,0) | ------ | 0. |
| f(0,1,1) | ------ | 0. |
| f (1,0,0) | ------ | 0. |
| f(1,0,1) | ------ | 0. |
| f(1,1,0) | ------ | 0. |
| f(1,1,1) | ------ | 0. |

Now the binary values of the corresponding positional number in the matrix are

```
    6  -----  1 1 0
    4  -----  1 0 0
    4  -----  1 0 0
         ---------------
              0 0 1
```
So the decimal value of 001 is 1. In this way the total posional values are calculated in the matrix

*A. Observation*
We have observed the matrix and also found some interesting fractal pattern in specific level wise(level 0,1,2…,). We also consider the level of the Boolean function upto 255 from 0.and consider the level in a signifant of manner ,e.g. from level 0 to 1, from level 2 to 3,from level 4 to 7,from level 8 to 15, from level 16 to 31 , from level 32 to 63, from level 64 to 127, from level 128 to 255.
1. From level (Z) 0 to 1 :- In this two level if we consider a 2x2 square matrix the following result is obtained:-

At level 0 a square matrix is found which is formed by only 0 entities and at level 1 a square matrix is also found formed by 0 and 1 entities.

The basic structure of the pattern of the square matrix at level 0 is below:-

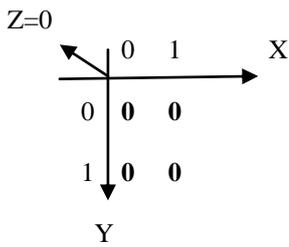

Figure-2    Formation of Euclidean geometry by the element 0

In this square matrix all combination is formed by 0
The basic structure of the pattern of the square matrix in level (Z) 1 as bellow—

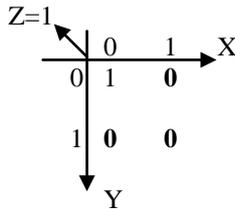

Figure-3    Formation of Fractal by the element 0

This is also a square matrix formed by 1 and 0. The pattern which is formed by 0 is symmetric to a right angle triangle

In this level(z=1) if we calculate a 4x4 matrix, then we can not get fractal pattern formed by matrix entities .of 0. The pattern is given below in the matrix:-

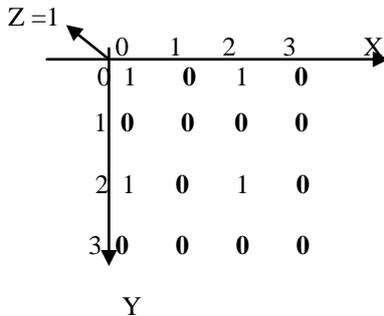

Figure-3    Formation of Euclidean geometry by the element 0

2. From level (Z) 2 to 3 :- In this two levels if we consider 4x4 square matrix, the same pattern which is formed by the 0 entities of the square matrix of 2x2 order at level 1 is repeated and regenerated .The basic pattern is here also fixed. The pattern is generated below:-

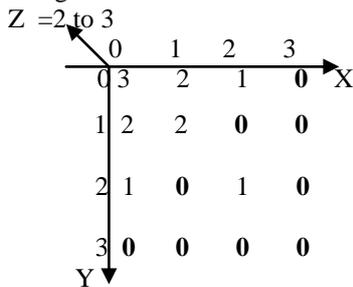

Figure-4    Formation of Fractal by the element 0

In this level if we also calculate 8x8 matrix, then we can not get the fractal pattern formed by 0 of the matrix . The pattern is given below:-

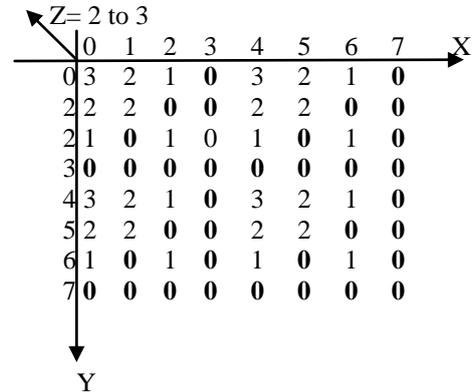

Figure-5    Formation of Euclidean geometry by the element 0

3. From level 4 to 7 :-   In this four levels if we consider a 8x8 square matrix, then at all levels (e.g. 4,5,6,7 )the same basic fractal pattern formed by the 0 entities of the square matrix of level 2 to 3 is repeated and also regenerated ,The pattern is given below:-

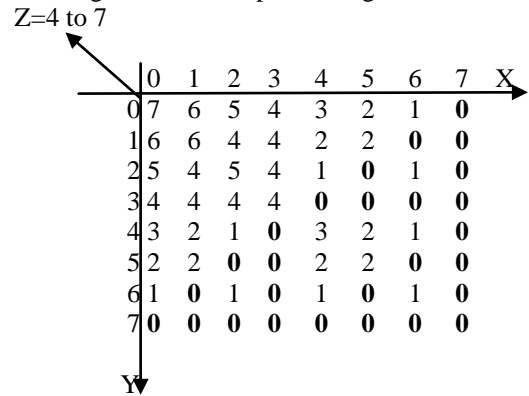

Figure -6    Formation of Fractal by the element 0

4.From level 8 to 15 :- In this levels if we consider 16x16 square matrix, at all levels ( e.g. 8,9,….15 levels) the same basic pattern formed by 0 entities of the square matrix of level 4 to 7 is repeated and regenerated upto level 15

Z=8 to 15

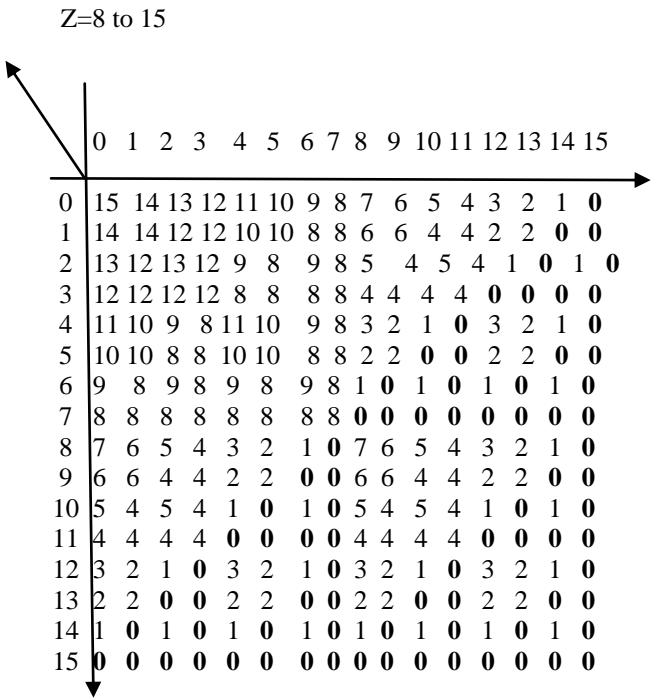

Y

Figure-7   Formation of Fractal by the element 0

5 From level 16 to 31 :- In this levels if we consider 32x32 square matrix, at all levels (e.g. 16,17,18…….31 levels) the same basic pattern formed by 0 entities of the square matrix of level  8 to 15 is repeated and also regenerated  upto level 31.

The same case is repeated and regenerated  in  also from level  32 to 63 if we consider 64x64 square matrix, and from 64 to 127   consider 128x128 square matrix ,and from 128 to 255 level  consider 255x255  square matrix .

*B.. Dimension of the pattern*

There are some level(Z) where we have got an Euclidean geometric figure(Figure-1,Figure-3,Figure-5…),so we can not calculate the dimension of that particular  pattern.

From level 1  to 255 if we calculate a particular square matrix, then the   dimension  of the pattern formed only by the entities  0  is fixed  and it is  D= log3/log2,
That is 1.5849 which is a Fractal.

*C. Analysis of the matrix*
We have analyzed the matrix and constructed a rule to get the corresponding level in which the fractal pattern belongs ..
1. From level 1 if we consider 2x2 matrix then we get the expected pattern of Fractals.
2. From level  2 to 3  if we consider 4x4 matrix then we get the  expected  pattern of Fractals.
3. Level 4 to 7 if we consider 8x8 matrix then also get  the expected pattern of Fractals.
4. Level 8 to 15  and 16 to 31 and so on upto level 255 we get the expected pattern of fractals if we consider the order of matrix nXn,
    Where n= the highest no. of  level +1.

Illustratati
Level 1
if we consider  the 1+1 =2 order matrix ,that is 2x2 matrix then we get the expected  self symmetric pattern of Fractals.

Level  2 to 3
If we consider the 2+2=4 order matrix ,that is 4x4  matrix then we get the same result.
In the same iteration  process  all levels satisfy the condition as below:-

Level    $2^0$ to $(2^{(0+1)} - 1)$    ----calculate   $2^{(0+1)} \times 2^{(0+1)}$ order matrix.
Level    $2^1$ to $(2^{(1+1)} - 1)$ ----- calculate   $2^{(1+1)} \times 2^{(1+1)}$ order matrix.
Level    $2^2$ to $(2^{(2+1)} - 1)$ ----- calculate   $2^{(2+1)} \times 2^{(2+1)}$ order matrix
Level    $2^3$ to $(2^{(3+1)} - 1)$ ------calculate  $2^{(3+1)} \times 2^{(3+1)}$ order matrix.
Level    $2^4$ to $(2^{(4+1)}-1)$ ------calculate  $2^{(4+1)} \times 2^{(4+1)}$ order matrix.
……………………………………………………………………
……..
……………………………………………………………………
……...
……………………………………………………………………
………
……………………………………………………………………
…………
By method of Induction
We can get

Level  (Z variable in the square matrix) $2^n$ to $(2^{(n+1)} - 1)$ ------- calculate   $2^{(n+1)} \times 2^{(n+1)}$ order matrix    to get the expected original pattern of fractals.
All levels  L $\in$ ( $2^n$ to $(2^{(n+1)} - 1)$ ).
Where n $\in$ Z+ ( set of all positive integer.)

V. Application

The application of these fractal patterns at different matrix dimension is towards the efficient as well as cost effective design of Wireless CDMA Networks..
Our aim of the implementation is to share small resources among the vast wireless CDMA network.

Let us take a small example;
our basic pattern comes like :-

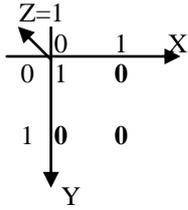

Figure-8   Basic Fractal pattern by the element 0 in 2x2 square matrix.

Suppose in this 2 x 2 matrix 4 blocks represent 4 resources and they access the wireless network. Now to provide the web service we need to have 4 active wireless web ports, now in our technique 4 ports are needed but we have to make active only 3 ports at an instant time and there is a certain switching devices which virtually rotates the apparent position of active wireless port of a certain clock speed in a specific direction. So, according to these concepts the signal that is sent to each user is same in respect of signal measurement and the clock speed is so high that it is seemed that the signal is continuous to the users. Now the quantities measure in respect of resource saving comes out in a very significant way. This process is like that whether we need to active 4 ports all times in a normal network structure, here just we need to active only 3 ports at all times and one port should stand by in its position. By some mathematical calculation we can say that our plan is $\{(1/4)*100\}\% = 25\%$ efficient to save the resource for this basic pattern of Fractal.

A typical diagram shows below to represent the Fractals in Wireless CDMA Network :-

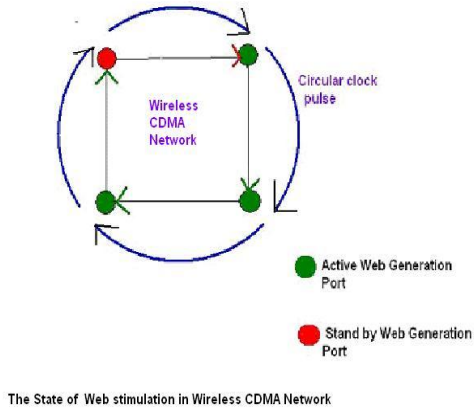

Figure -9

If we consider our general formula that is Level $2^n$ to $(2^{(n+1)} - 1)$ -------calculate $2^{(n+1)} \times 2^{(n+1)}$ order matrix to get the expected original pattern of fractals.

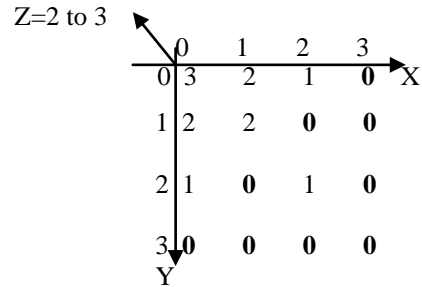

Figure-10   Formation of Fractal by the entities 0 in 4x4 Square matrix.

As the rule if we consider 4x4 matrix for level 3 the same pattern of the fractals is repeated and regenerated.
In this case we also apply the previous concept, if the total pattern rotates in a certain clock speed x in a certain direction; then the elementary pattern that is depicted below have to rotate in 4x speed at the same direction. Here the prefix 4 is to be added because after changing the initial position again returning to the initial position one pattern has to change position 4 times. By the help of simple arithmetic as the previous one we are able to calculate the % of efficient shave of the recourse;
$\{(7/16)*100\}\% = 43.75\%$
It seems that our efficiency has grown up.
Then if we calculate as our proposed formula 8x8 matrix at level 7 then the pattern is repeated as well as regenerated.

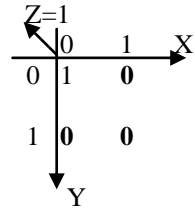

Figure -11   Elementary Fractal pattern by the element 0 in 2x2 Square matrix.

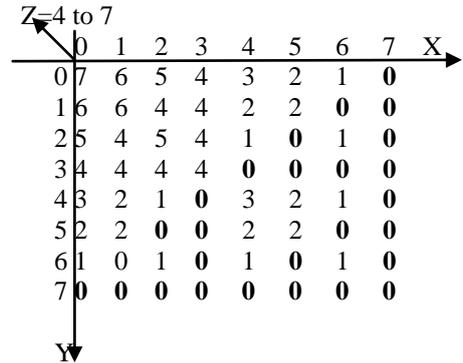

Figure-12   Fractal pattern by the element 0 in 8x8 Square matrix

Apply and approach same as the previous one. Now in this case if the total pattern rotates in certain clock speed x in certain direction then the corresponding elementary pattern level 1

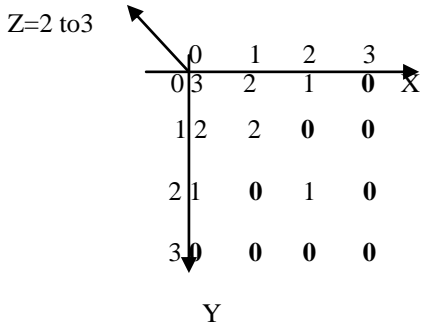

Figure-13    Fractal pattern by the element 0

have to switch in 4x speed   in same direction and the corresponding elementary pattern  level 2

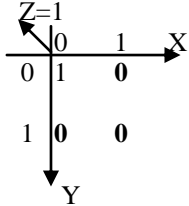

Figure- 14   Elementary Fractal pattern by the element 0

have to rotate in $4^2$x speed. Same simple clock logic as the previous one. From the calculation as the previous one we are able to calculate the % of efficiency to optimizes the resources.

$\{(37/64)*100\}\% = 57.8125\%$

It shows  that our efficiency has grown up gradually. Then the same  occurrence repeated of the matrix  arrangement at level 7 as   the pattern is generated in case of  16 x 16 matrix and also the
% of efficiency for saving the resources  increases;

$\{(172/256)*100\}\%=67.18\%$

Same pattern is regenerated for consideration of   32 x 32 matrix  at level  16,  the % of efficiency  here  is also save of the recourse and becomes;

$\{(781/1024)*100\}\%=76.26\%$

As the process goes on we can see that the efficiency of our implementation increases;
So we can say that our proposed  concept  is more  efficient for the large coverage area of a wireless  CDMA network.
In all above case we have to increase the speed elementary pattern level value ,( the level value may vary from 1 ,2,3,4,,5, ,n; where n is any natural positive number.) at the order of 4 level value . So according to this if the total pattern rotates at certain clock speed x ; then the speed of the then the elementary pattern at different level rotates according to the multiple of their corresponding level value . So, from the above discussion we can conclude that the speed of the particular elementary pattern at the particular level depends on the level value and that can be expressed as [ $4^{\text{level value}} * x$ ].

VI. Comparison of our CDMA system with existing CDMA system

*A. Advantages*

There are mainly three advantages of our proposed CDMA system over the existing CDMA system. First of all number of resources are getting reduced at any instant time without affecting the network service to the station . Secondly as the number of resources  getting reduced, so the cost of network service is also getting reduced. Lastly at any instant time the number of resources  will  remain standby in our CDMA system, so the energy will also save.

*B. Limitations*

There is a limitation to construct this CDMA network on the basis of fractal pattern, To get the fractal pattern the number of resources should be in our proposed manner, that means if we imagine the resources of the network as in the position of a matrix entities, then all the resources will be reconstructed according to the  position of the entities of a square matrix. If the above conditions will satisfy then, it is possible to construct such network model on the basis of fractal pattern.

VII. Conclusion and Further Research

In this paper we have used three variable Boolean function 3 in formation of fractal pattern   and also completing the analysis we can say  at  level  $2^n$  to $(2^{(n+1)} -1)$ of a level of a square matrix    if we consider the  $2^{(n+1)} \times 2^{(n+1)}$   square matrix the fractal  pattern of  the dimension same as the dimension of sierpinski triangle that means  the dimension will be 1.5849 is produced. From this pattern model we can designing  a new area of  Wireless CDMA Network   that can be used for a efficient and effective resources   saving  Wireless CDMA Communication  .
.
In further case if we try to implement the a new Wireless LAN technology ,it is possible to construct from our  research .